\documentclass[twocolumn]{autart}
\usepackage[authoryear]{natbib}

\usepackage{graphics}
\usepackage{graphicx}
\usepackage{epsfig}
\usepackage{color}
\usepackage{subfigure}

\usepackage{amssymb}
\usepackage{amsmath}
\usepackage{amsfonts}
\usepackage{stmaryrd}
\usepackage{txfonts}
\usepackage{mathrsfs}
\usepackage{latexsym, bm}
\usepackage{xcolor,stfloats}
\usepackage{lipsum}%

\usepackage{indentfirst}
\usepackage{amsmath,amssymb,bm,bbm}
\usepackage{graphicx}
\usepackage{latexsym}
\usepackage{amssymb}

\def\QEDclosed{\mbox{\rule[0pt]{1.3ex}{1.3ex}}} 

\def\QED{\QEDclosed} 
\def\pf{\noindent{\bf Proof. }}
\def\endpf{\hspace*{\fill}~\QED\par\endtrivlist\unskip}

\newtheorem{theorem}{Theorem}

\newtheorem{lemma}[theorem]{Lemma}
\newtheorem{definition}[theorem]{Definition}
\newtheorem{remark}{Remark}

 \DeclareMathOperator{\diag}{diag}

\begin{document}

\begin{frontmatter}

\title{Containment control of multi-agent systems with measurement noises\thanksref{label1}}
\thanks[label1]{Yuanshi Zheng's work is supported by the National Natural Science Foundation
of China under grants 61375120 and 61304160 and the Fundamental Research Funds
for the Central Universities under grant K5051304049. Tao Li's work is supported by the National Natural Science Foundation
of China under grant 61370030. Long Wang's work is supported by the 973 Program under grant 2012CB821203 and the National Natural Science Foundation of China under grants 61020106005 and 61375120. This paper was not presented at any IFAC meeting.
}

\author[xidian]{Yuanshi Zheng }, \ead{zhengyuanshi2005@163.com}
\author[shanghai]{Tao Li }, \ead{sixumuzi@shu.edu.cn}
\author[beida]{Long Wang \corauthref{cor1}}  \ead{longwang@pku.edu.cn}
\corauth[cor1]{Corresponding author : Long Wang }
\address[xidian]{ Center for Complex Systems, School of Mechano-electronic Engineering, Xidian University, Xi'an 710071, China}
\address[shanghai]{ Department of Automation, School of Mechatronic Engineering and Automation, Shanghai University, Shanghai 200072, China}
\address[beida]{Center for Systems and Control, College of Engineering, Peking
University, Beijing 100871, China}

\begin{abstract}
In this paper, containment control of multi-agent systems with measurement noises is studied under directed networks. When the leaders are stationary, a stochastic approximation type protocol is employed to solve the containment control of multi-agent systems. By using stochastic analysis tools and algebraic graph theory, some necessary and sufficient criteria are established to ensure the followers converge to the convex hull spanned by the leaders in the sense of mean square and probability $1$. When the leasers are dynamic, a stochastic approximation type protocol with distributed estimators is developed and necessary and sufficient conditions are also obtained for solving the containment control problem. Simulations are provided to illustrate the effectiveness of the theoretical results.
\end{abstract}
\end{frontmatter}
\section{\bf Introduction}\label{s-introduction}

In recent years, cooperative control of multi-agent systems has attracted considerable attention in the control community, thanks to the broad applications in various areas, such as biology (e.g., flocking of birds \citep{Saber06}), social sciences (e.g., evolution of language \citep{cucker04}) and engineering (e.g., formation control of UAVs \citep{xiao09}), etc. Like consensus problem \citep{Olfati-Saber07,Xie07,zheng11-2}, containment control is also regarded as a fundamental problem in the cooperative control of multi-agent systems. Containment control means that a group of agents converge to the convex hull which is spanned by other group of agents. Additional details related to the containment control problem may be obtained from references therein.

The consensus problem was primarily considered without leader \citep{jadbabaie03,saber04,ren05}. However, a group of agents might have one or multiple leaders in the agent network. For the single leader, the followers will converge to the state of leader, which is called the tracking control or leader-following consensus problem. In \cite{hong06}, tracking control problem for multi-agent systems was studied with an active leader under time-varying undirected topology. In \cite{ni2010}, leader-following consensus of high-order multi-agent systems was considered with fixed and switching topologies. \cite{zheng12-3} investigated the consensus of heterogeneous multi-agent systems with a time-varying group reference velocity. If the multi-agent system has multiple leaders, the objective is to drive the followers into the convex hull spanned by the leaders, which is called the containment control problem. Motivated by the numerous natural phenomena and applications in engineering, containment control has been considered by lots of researchers in recent years.  In \cite{ji08}, a Stop-Go control policy was proposed to drive a collection of the followers to a given target destination which is spanned by the leaders. In \cite{meng10}, finite-time attitude containment control was considered for multiple rigid bodies under undirected topology. \cite{Notarstefano11} considered containment control of first-order multi-agent systems with switching topologies. Containment control of second-order multi-agent systems with multiple stationary and dynamic leaders was studied under fixed and switching topologies in \cite{cao11}. \cite{lou12} investigated the containment control of second-order multi-agent systems with random switching topologies. \cite{liu12Auto} gave some necessary and sufficient conditions for the containment control of multi-agent systems with multiple stationary and dynamic leaders under directed networks. In \cite{Zheng14}, the containment control of heterogeneous multi-agent systems was also discussed under directed networks.

It should be noted that all the aforementioned results were assumed that the communication channel is ideal. However, real networks are often unideal because of uncertain communication environment. Thus, it is natural to consider the noise in the distributed protocols for cooperative control of multi-agent systems. Huang and Manton proposed a a stochastic approximation type protocol for discrete-time first-order multi-agent systems with
measurement noises \citep{Huang09} and proved the mean square consensus and strong consensus under randomly varying topologies \citep{Huang10}. \cite{li10} studied the stochastic approximation type protocol under time-varying topologies and proved that mean square and almost sure consensus can be ensured if the network switches among balanced and jointly-containing-spanning-tree graphs. In \cite{li09}, the continuous-time stochastic approximation type consensus was studied and necessary and sufficient conditions on the control gain function were given. \cite{wangbingchang09} extended the results in \cite{li09} to study the mean square consensus and strong consensus of continuous-time first-order multi-agent systems under general fixed directed graph. The result of \cite{li09} was also extended to the second-order case with local feedback in \cite{cheng11}. \cite{zheng11-1} studied the finite time consensus of stochastic multi-agent systems. It is noted that the tracking control and containment control of multi-agent systems with measurement noises only receives a little attention. In \cite{hu10}, tracking control of multi-agent systems was considered with an active leader and measurement noises under directed topology and some sufficient conditions were obtained for solving the mean square consensus. \cite{tang2012} studied the containment control of discrete-time multi-agent systems with multiple stationary leaders and noisy measurements and gave some sufficient conditions for solving the consensus almost surely and in mean square.

Inspired by the recent developments in containment control and stochastic approximation type consensus protocols, we investigate the containment control of multi-agent systems with measurement noises under the fixed directed topology in this paper. Different from \cite{tang2012}, we consider continuous-time multi-agent systems with multiple stationary and dynamic leaders, respectively. Containment control of continuous-time multi-agent systems for noise-free cases has been widely studied. For example, \cite{liu12Auto} investigated the containment control of multi-agent systems without noise and obtained some necessary and sufficient conditions for solving the containment control with multiple stationary and dynamic leaders under the directed topology. Due to the measurement noises, the distributed protocol should be designed with the time-varying gains $a(t)$ and the closed-loop system becomes a time-varying stochastic differential equation. The Laplacian matrix is asymmetric in a directed graph.  All of these make difficulties and challenges for the convergence analysis of containment control problem. The main contribution of this paper is twofold. Firstly, we employ a stochastic approximation type protocol for the multi-agent system with multiple stationary leaders. By using stochastic Lyapunov theory, algebraic graph theory and Doob's martingale convergence theorem, some necessary and sufficient conditions are presented for solving the containment control in the sense of mean square and probability $1$ under the directed topology. Secondly, a stochastic approximation type protocol with distributed estimators is developed for the multi-agent system with multiple dynamic leaders. Similar to the case of  multiple stationary leaders, we obtain some necessary and sufficient conditions for solving the containment control in the sense of mean square and probability $1$.

The rest of this paper is organized as follows. In Section \ref{s-preliminaries}, we present some notions in graph theory and assemble some key lemmas. We provide convergence analysis for containment control of multi-agent systems with multiple stationary leaders in Section \ref{s-stationary} and multiple dynamic leaders in Section \ref{s-dynamic}. In Section \ref{s-Simulations}, numerical simulations are given to illustrate the effectiveness of the theoretical results. In Section \ref{s-Conclusion}, Some conclusions are drawn.

The following notations will be used throughout this paper: $\mathbb{R}$ denotes the set of real  number, $\mathbb{R}^N$ denotes the $N-$dimensional real vector space. $\mathcal{I}_{n}=\{1,2, \dots,\ n\}$. For a given vector or matrix $A,$ $A^{T}$ denotes its transpose, $tr(A)$ denotes its trace when $A$ is square. $\diag\{a_{1},a_{2} \cdots,\ a_{n}\}$ defines a diagonal matrix with diagonal elements being $a_{1}, a_{2} \cdots,\ a_{n}$. For a given random variable $X$, $E(X)$ denotes its mathematical expectation.
$\mathbf{1}_{n}$  is a vector with elements being all ones. $I_{n}$ is the $n\times n$ identity matrix.
$0$ $(0_{m\times n})$ denotes an all-zero vector or matrix with compatible dimension (dimension $m\times n$). $A \otimes B$ denotes the Kronecker product of matrices $A$ and $B$.


\section{\bf Preliminaries}\label{s-preliminaries}

In this section, we fist introduce some basic concepts and results about graph theory. For more details, please refer to \cite{Godsil01}. Then, some lemmas and definitions are given which will be used in this paper.

A weighted directed graph $\mathscr{G}(\mathscr{A})=(\mathscr{V},\mathscr{E},\mathscr{A})$ of order $n$
consists of a vertex set $\mathscr{V}=\{s_{1}, s_{2}, \cdots, s_{n}\}$,  an edge set
$\mathscr{E}=\{e_{ij}=(s_{i}, s_{j})\}\subset \mathscr{V}\times \mathscr{V}$ and a nonnegative
matrix $\mathscr{A}=[a_{ij}]_{n\times n}$.  A directed path between two distinct vertices $s_{i}$ and $s_{j}$ is
a finite ordered sequence of distinct edges of $\mathscr{G}$ with the form $(s_{i}, s_{k_{1}}), (s_{k_{1}}, s_{k_{2}}), \cdots, (s_{k_{l}}, s_{j})$.
A directed tree is a directed graph, where there exists a vertex called the root such that there exists a unique directed path from this vertex to every other vertex. A directed forest is a directed graph which consists of two or more directed trees satisfying
none of two have a common vertex. A directed spanning forest of the directed graph $\mathscr{G}$ is a directed forest which consists of all the vertices and
some edges in $\mathscr{G}$. The degree matrix
$\mathscr{D}=[d_{ij}]_{n\times n}$ is a diagonal matrix with $d_{ii}=\sum_{j:s_j\in \mathscr{N}_{i}} a_{ij}$ and
the Laplacian matrix of the graph is defined as $\mathscr{L}=[l_{ij}]_{n \times n}=\mathscr{D}-\mathscr{A}.$

For an $n$-agent system, an agent is called a leader if
the agent has no neighbor, and an agent is called a follower if the agent has at least one neighbor. Suppose that the multi-agent system has $m$ leaders and $n-m$ followers. Denote the set of leaders as $\mathscr{R}$ and the set of followers as $\mathscr{F}$, respectively. Then, $\mathscr{L}$ can be partitioned as
$\mathscr{L}=\begin{pmatrix} \mathscr{L}_{\mathscr{F}\mathscr{F}} & \mathscr{L}_{\mathscr{F}\mathscr{R}} \\ 0_{m\times(n-m)}& 0_{m\times m} \end{pmatrix}$.

\begin{lemma}(\cite{liu12Auto})\label{lemma-1}
$\mathscr{L}_{\mathscr{F}\mathscr{F}}$ is invertible if and only if the directed graph $\mathscr{G}$ has a directed spanning forest with the all the root nodes being the leaders.
\end{lemma}

In fact, $-\mathscr{L}_{\mathscr{F}\mathscr{F}}$ is Hurwitz. Thus, there exists a positive definite matrix $P$ such that $P\mathscr{L}_{\mathscr{F}\mathscr{F}}+\mathscr{L}_{\mathscr{F}\mathscr{F}}^{T}P=I_{n-m}$.

\begin{lemma}(Doob's martingale convergence theorem \cite{mao1997})\label{lemma-2}
Let $\left\{M_{t}\right\}_{t\geq0}$ be a real-valued right-continuous supermartingale. If
$$\sup_{0\leq t<\infty}E M_{t}^{-}<\infty,$$
then $M_{t}$ converges almost surely to a random variable $M_{\infty}$. In particular, this holds if $M_{t}$ is nonnegative.
\end{lemma}

\begin{definition}\label{def-convex} (\cite{Rockafellar72})
A subset $\mathcal{C}$ of $\mathbb{R}^{m}$ is said to be convex if $(1-\lambda)x+\lambda y\in\mathcal{C}$ whenever
$x\in\mathcal{C}$, $y\in\mathcal{C}$ and $0<\lambda<1$. The convex hull of a finite set of
points $X=\{x_{1}, \ldots , x_{n}\}$ in $\mathbb{R}^{m}$ is the minimal convex set containing all
points in $X$, denoted by $Co\{X\}$. Particularly, $Co\{X\}\triangleq\{\sum_{i=1}^{n}\alpha_{i}x_{i}|x_{i}\in X, \alpha_{i}\geq 0, \sum_{i=1}^{n}\alpha_{i}=1\}$.
\end{definition}

\begin{definition}\label{def-mean square}
We say that the multi-agent system solves the containment control in mean square if for any initial conditions, we have $\lim_{t\rightarrow \infty}E\|x_i(t)-Co(\mathscr{R})\|^{2}=0$, for $i\in\mathscr{F}$.
\end{definition}

\begin{definition}\label{def-almost surely}
We say that the multi-agent system solves the containment control almost surely if for any initial conditions, we have $\lim_{t\rightarrow \infty}\|x_i(t)-Co(\mathscr{R})\|=0$ almost surely (a.s.), for $i\in\mathscr{F}$.
\end{definition}

\section{\bf Containment control with multiple stationary leaders}\label{s-stationary}
In this section, we consider a multi-agent system which is consisted of $n$ first-order integrator agent with dynamics
\begin{equation}\label{m-first}
   \dot{x}_{i}(t)=u_{i}(t),  ~~~~i\in\mathcal{I}_{n},
  \end{equation}
where $x_{i}\in \mathbb{R}^{N}$ and $u_{i}\in \mathbb{R}^{N}$ are the position and control input of agent $i$, respectively.

Due to the existence of noises in communication channel, we give a protocol with measurement noises for multi-agent system (\ref{m-first}) as follows:
\begin{equation}\label{p-first}
   u_{i}(t)=\left\{
   \begin{aligned}
   &a(t)\sum_{j\in \mathscr{F}\cup\mathscr{R}}a_{ij}(y_{ji}(t)-x_i(t)), & i \in \mathscr{F},\\
   &0,   &i\in \mathscr{R},\\
   \end{aligned}
   \right.
  \end{equation}
where $y_{ji}(t)=x_{j}(t)+\sigma_{ji}\eta_{ji}(t)$ denotes the measurement of the $j$th agent's position,
$\{\eta_{ji}(t)\in \mathbb{R}^{N}, i,j=1, 2, \ldots, n\}$ are independent standard white noises, $\sigma_{ji} \geq 0$ is the noise intensity, $\mathscr{A}=[a_{ij}]_{n\times n}$ is the weighted adjacency matrix, $a(\cdot): [0, \infty)\rightarrow [0, \infty)$ is called the gain function.

To get the main results, we give the following assumptions:\\
(A1) For any given follower $j\in\mathscr{F}$, there is a leader $i\in\mathscr{R}$, such that there exists a directed path from $i$ to $j$.\\
(A2) $\int_{0}^{\infty}a(s)ds=\infty$.\\
(A3) $\int_{0}^{\infty}a^{2}(s)ds<\infty$.\\
(A4) $\lim_{t\rightarrow\infty} a(t)=0$.

\begin{remark}\label{remark-leader}
In this paper, we consider the multi-agent system with multiple leaders and followers. Thus,  it is easy to know that Assumption (A1) holds if and only if the directed graph $\mathscr{G}(\mathscr{A})$ has a directed spanning forest with all the root nodes being the leaders.
\end{remark}

\begin{theorem}\label{T-first-mean}
Consider a directed fixed network $\mathscr{G}(\mathscr{A})$. Assume that (A3) holds. Then, the multi-agent system (\ref{m-first}) with protocol (\ref{p-first}) solves the containment control in mean square if and only if (A1)--(A2) hold.
\end{theorem}
{\bf Proof.} Sufficiency. The multi-agent system (\ref{m-first}) with protocol (\ref{p-first}) can be rewritten in the form of the It$\hat{o}$ stochastic differential equation
\begin{equation}\label{m-first-p}
\left\{
   \begin{aligned}
  &dx_{\mathscr{F}}(t)=-a(t)\left(\mathscr{L}_{\mathscr{F}\mathscr{F}}\otimes I_{N}\right)x_{\mathscr{F}}(t)dt\\
   &~~~~~~-a(t)\left(\mathscr{L}_{\mathscr{F}\mathscr{R}}\otimes I_{N}\right)x_{\mathscr{R}}(t)dt+a(t)\left(\Omega\otimes I_{N}\right)dW(t),\\
  &\frac{dx_{\mathscr{R}}(t)}{dt}=0,
  \end{aligned}
   \right.
  \end{equation}
where $\Omega=\diag\left(\sqrt{\sum_{j=1}^{n}\left(a_{1j}\sigma_{j1}\right)^{2}}, \cdots, \sqrt{\sum_{j=1}^{n}\left(a_{nj}\sigma_{jn}\right)^{2}}\right)$,  $W(t)=\left(W_{1}(t), W_{2}(t), \cdots, W_{n}(t)\right)$ is an $n$ dimensional standard Brownian motion.

Because Assumption (A1) holds, we know that $\mathscr{L}_{\mathscr{F}\mathscr{F}}$ is invertible from Lemma \ref{lemma-1}. 
Let $\delta(t)=x_{\mathscr{F}}(t)+\left(\left(\mathscr{L}_{\mathscr{F}\mathscr{F}}^{-1}\mathscr{L}_{\mathscr{F}\mathscr{R}}\right)\otimes\right.$ $\left.I_{N}\right)x_{\mathscr{R}}(t)$. Then, we have
\begin{equation}\label{m-delta}
   d\delta(t)=-a(t)\left(\mathscr{L}_{\mathscr{F}\mathscr{F}}\otimes I_{N}\right)\delta(t)dt+a(t)\left(\Omega\otimes I_{N}\right)dW(t).
  \end{equation}
 Let $V(t)=\delta^{T}(t)\left(P\otimes I_{N}\right)\delta(t)$. Differentiating $V(t)$, yields that
\begin{equation}\label{dV(t)}
\begin{aligned}
    dV(t)&=-a(t)\delta^{T}(t)\delta(t)dt+2a(t)\delta^{T}(t)\left(P\Omega\otimes I_{N}\right)dW(t)\\
    &~~~~~~+a^{2}(t)Tr\left(P\Omega \Omega^{T}\otimes I_{N}\right)dt\\
    &\leq-\frac{1}{\lambda_{max}}a(t)V(t)dt+2a(t)\delta^{T}(t)\left(P\Omega\otimes I_{N}\right)dW(t)\\
     &~~~~~~+a^{2}(t)Tr\left(P\Omega \Omega^{T}\otimes I_{N}\right)dt,
\end{aligned}
\end{equation}
where $\lambda_{max}$ is the maximum eigenvalue of $P\otimes I_{N}$. By known results in \cite{yong1999}, for any $T>0$, there exists a constant $C_{T}$, such that
\begin{equation}\label{E(delta)}
   E \sup _{0\leq t\leq T}\|\delta(t)\|^{2}\leq C_{T}\left(1+E\|\delta(0)\|^{2}\right)<\infty,
\end{equation}
which implies that $\int_{0}^{t}2a(s)\delta^{T}(s)\left(P\Omega\otimes I_{N}\right)dW(s)$ is a martingale, i.e.,

\begin{equation}\label{martingale}
 E \int_{0}^{t}2a(s)\delta^{T}(s)\left(P\Omega\otimes I_{N}\right)dW(s)=0.
\end{equation}

Therefore,
\[
\begin{aligned}
  &E V(t)\leq V(0)\\
  &~~~~~~~+\int_{0}^{t}\left[-\frac{1}{\lambda_{max}}a(s)E V(s)+a^{2}(s)Tr\left(P\Omega \Omega^{T}\otimes I_{N}\right)\right]ds.
\end{aligned}
\]
Owing to $\int_{0}^{\infty}a^{2}(s)ds<\infty$, we have $\int_{0}^{t}a^{2}(s)Tr\left(P\Omega \Omega^{T}\otimes I_{N}\right)ds$ $<\infty$. Let $C_{1}= V(0)+\int_{0}^{t}a^{2}(s)Tr\left(P\Omega \Omega^{T}\otimes I_{N}\right)ds<\infty$, we get
\[
\begin{aligned}
  E V(t)\leq \int_{0}^{t}-\frac{1}{\lambda_{max}}a(s)E V(s)ds+C_{1}.
\end{aligned}
\]
By comparison principle and Assumption (A2), we have
\[
\begin{aligned}
  E V(t)\leq C_{1}e^{-\frac{1}{\lambda_{max}}\int_{0}^{t}a(s)ds}\rightarrow 0,
\end{aligned}
\]
as $t\rightarrow\infty$, which implies that $\lim_{t\rightarrow0}E\|\delta(t)\|^{2}=0$. Thus,
\[
\begin{aligned}
  \lim_{t\rightarrow0}E\|x_{\mathscr{F}}(t)+\left(\left(\mathscr{L}_{\mathscr{F}\mathscr{F}}^{-1}\mathscr{L}_{\mathscr{F}\mathscr{R}}\right)\otimes I_{N}\right)x_{\mathscr{R}}(0)\|^{2}=0.
\end{aligned}
\]

From the definition of $\mathscr{L}$, we can obtain that \\ $\left(-\left(\left(\mathscr{L}_{\mathscr{F}\mathscr{F}}^{-1}\mathscr{L}_{\mathscr{F}\mathscr{R}}\right)\otimes I_{N}\right)\right)\left(\mathbf{1}_{m}\otimes\mathbf{1}_{N}\right)=\mathbf{1}_{n-m}\otimes\mathbf{1}_{N}$. Hence, the multi-agent system (\ref{m-first-p}) solves the containment control in mean square and the final position of the followers are given by $-\left(\left(\mathscr{L}_{\mathscr{F}\mathscr{F}}^{-1}\mathscr{L}_{\mathscr{F}\mathscr{R}}\right)\otimes I_{N}\right)x_{\mathscr{R}}(0)$.

Necessity. Firstly, we prove the necessity of (A1). Similar to the proof of Theorem 1 in \cite{liu12Auto}, we use the reduction to absurdity. If there exists at least one follower such that it does not have a directed path from the leader to this follower, the position of this follower is independent of the position of the leaders, which implies that the containment control problem can not be solved.

Secondly, we prove the necessity of (A2). From (\ref{dV(t)}) and (\ref{martingale}), we can get
\[
\begin{aligned}
  &EV(t)-EV(0)\\
  &\geq -\frac{1}{\lambda_{min}}\int_{0}^{t}a(s)E V(s)ds+\int_{0}^{t}a^{2}(s)Tr\left(P\Omega \Omega^{T}\otimes I_{N}\right)ds\\
  &\geq -\frac{1}{\lambda_{min}}\int_{0}^{t}a(s)E V(s)ds,
\end{aligned}
\]
where $\lambda_{min}$ is the minimum eigenvalue of $P\otimes I_{N}$. Thus,
\[
\begin{aligned}
  EV(t)\geq EV(0)e^{-\frac{1}{\lambda_{min}}\int_{0}^{t}a(s)ds}.
\end{aligned}
\]
In view of $EV(0)>0$, $\lambda_{min}>0$ and $EV(t)\rightarrow 0$, we have $\int_{0}^{\infty}a(s)ds=\infty$. $\blacksquare$

In fact, Assumption (A3) is unnecessary. Similar to Theorem $4$ in \cite{wangbingchang09}, we have the result as follows.

\begin{theorem}\label{T-first-mean-2}
Consider a directed fixed network $\mathscr{G}(\mathscr{A})$. Assume that (A4) holds. Then, the multi-agent system (\ref{m-first}) with protocol (\ref{p-first}) solves the containment control in mean square if and only if (A1)--(A2) hold.
\end{theorem}
{\bf Proof.} Sufficiency. From (\ref{dV(t)}), we have
\[
\begin{aligned}
   EV(t)\leq V(0)+\int_{0}^{t}\left[-\frac{1}{\lambda_{max}}a(s)E V(s)+a^{2}(s)Tr\left(P\Omega \Omega^{T}\otimes I_{N}\right)\right]ds.
\end{aligned}
\]
Using comparison principle, we get
\[
\begin{aligned}
   &EV(t)\leq V(0)e^{-\frac{1}{\lambda_{max}}\int_{0}^{t}a(s)ds}\\
   &~~~~~~~~~~+Tr\left(P\Omega \Omega^{T}\otimes I_{N}\right)\int_{0}^{t}e^{-\frac{1}{\lambda_{max}}\int_{s}^{t}a(u)du}a^2(s)ds.
\end{aligned}
\]
Owing to $\int_{0}^{\infty}a(s)ds=\infty$, we have $\lim_{t\rightarrow\infty}V(0)e^{-\frac{1}{\lambda_{max}}\int_{0}^{t}a(s)ds}=0$.

Due to (A4), there exists $C_{2}$ such that $a(t)<C_{2}$ for $t\geq0$, and for any $\varepsilon>0$, there exists $t_{0}>0$ such that $a(t)<\varepsilon$ for $t>t_{0}$. Thus,
\[
\begin{aligned}
  &\int_{0}^{t}e^{-\frac{1}{\lambda_{max}}\int_{s}^{t}a(u)du}a^2(s)ds=\left( \int_{0}^{t_{0}}+ \int_{t_{0}}^{t}\right)e^{-\frac{1}{\lambda_{max}}\int_{s}^{t}a(u)du}a^2(s)ds\\
  &\leq C_{2}\lambda_{max}\int_{0}^{t_{0}}d\left(e^{-\frac{1}{\lambda_{max}}\int_{s}^{t}a(u)du}\right)+
  \lambda_{max}\varepsilon\int_{t_{0}}^{t}d\left(e^{-\frac{1}{\lambda_{max}}\int_{s}^{t}a(u)du}\right)\\
  &= C_{2}\lambda_{max}\left(e^{-\frac{1}{\lambda_{max}}\int_{t_{0}}^{t}a(u)du}-e^{-\frac{1}{\lambda_{max}}\int_{0}^{t}a(u)du}\right)+
  \lambda_{max}\varepsilon\left(1-e^{-\frac{1}{\lambda_{max}}\int_{t_{0}}^{t}a(u)du}\right)
\end{aligned}
\]
Because $\lim_{t\rightarrow\infty}e^{-\frac{1}{\lambda_{max}}\int_{t_{0}}^{t}a(u)du}=\lim_{t\rightarrow\infty}e^{-\frac{1}{\lambda_{max}}\int_{0}^{t}a(u)du}=0$, we have
\[
\begin{aligned}
  \lim_{t\rightarrow\infty}\int_{0}^{t}e^{-\frac{1}{\lambda_{max}}\int_{s}^{t}a(u)du}a^2(s)ds=0.
\end{aligned}
\]
Thus, similar to the proof of sufficiency in Theorem \ref{T-first-mean}, the multi-agent system (\ref{m-first}) with protocol (\ref{p-first}) solves the containment control in mean square.

The proof of necessity can be found in Theorem \ref{T-first-mean}.    $\blacksquare$

\begin{theorem}\label{T-first-almost}
Consider a directed fixed network $\mathscr{G}(\mathscr{A})$. Assume that (A3) holds. Then, the multi-agent system (\ref{m-first}) with protocol (\ref{p-first}) solves the containment control almost surely if and only if (A1)--(A2) hold.
\end{theorem}
{\bf Proof.} Sufficiency. According to (\ref{dV(t)}) in Theorem \ref{T-first-mean}, we have
\[
\begin{aligned}
    &V(t)=V(0)+\int_{0}^{t}a^{2}(s)Tr\left(P\Omega \Omega^{T}\otimes I_{N}\right)ds\\
    &-\int_{0}^{t}a(s)\delta^{T}(s)\delta(s)ds+\int_{0}^{t}2a(s)\delta^{T}(s)\left(P\Omega\otimes I_{N}\right)dW(s).
\end{aligned}
\]
Let $\tilde{V}(t)=V(t)-\int_{0}^{t}a^{2}(s)Tr\left(P\Omega \Omega^{T}\otimes I_{N}\right)ds$. Then, for any $r\leq t$,
\[
\begin{aligned}
   & E\left(\tilde{V}(t)-\tilde{V}(r)\mid\mathcal{F}_{r}\right)\\
       &=E\left(-\int_{r}^{t}a(s)\delta^{T}(s)\delta(s)ds\right.\\
       &~~~~\left.+\int_{r}^{t}2a(s)\delta^{T}(s)\left(P\Omega\otimes I_{N}\right)dW(s)\mid\mathcal{F}_{r}\right)\\
       &=E\left(-\int_{r}^{t}a(s)\delta^{T}(s)\delta(s)ds\mid\mathcal{F}_{r}\right)\leq 0,
\end{aligned}
\]
which implies that $\tilde{V}(t)$ is a supermartingale.

Owing to $V(t)\geq 0$ and $\int_{0}^{\infty}a^{2}(s)ds<\infty$, we have
\[
\begin{aligned}
  \sup_{t\geq0} E\left(\tilde{V}^{-}(t)\right)\leq \int_{0}^{\infty}a^{2}(s)Tr\left(P\Omega \Omega^{T}\otimes I_{N}\right)ds<\infty,
\end{aligned}
\]
where $\tilde{V}^{-}(t)=\max\left\{-\tilde{V}(t), 0\right\}$.

By Lemma \ref{lemma-2}, there exists $\tilde{V}^{\ast}$ such that $\tilde{V}(t)\rightarrow \tilde{V}^{\ast}$ a.s.. Owing to (A3), it implies that $V(t)$ is convergent almost surely.
From Theorem \ref{T-first-mean}, we know that $E V(t)\rightarrow 0$. Thus, we get $V(t)\rightarrow 0$ a.s., which implies $\lim_{t\rightarrow0}\delta(t)=0$ a.s.. Therefore, the multi-agent system (\ref{m-first}) with protocol (\ref{p-first}) solves the containment control almost surely and the final position of the followers are given by $-\left(\left(\mathscr{L}_{\mathscr{F}\mathscr{F}}^{-1}\mathscr{L}_{\mathscr{F}\mathscr{R}}\right)\otimes I_{N}\right)x_{\mathscr{R}}(0)$.

Necessity. The necessity of (A1) can be proved by the same argument as in the proof of the necessity in Theorem \ref{T-first-mean}.

Next, we prove the necessity of (A2). It is easy to know that $\Phi(t)=e^{-\int_{0}^{t}a(s)\left(\mathscr{L}_{\mathscr{F}\mathscr{F}}\otimes I_{N}\right)ds}$ is the fundamental solution of the equation $\dot{\delta}(t)=-a(t)\left(\mathscr{L}_{\mathscr{F}\mathscr{F}}\otimes I_{N}\right)\delta(t)$. By the variation-of-constants formula, the solution of equation (\ref{m-delta}) can be expressed as
\begin{equation}\label{delta}
   \delta(t)=\Phi(t)\left(\delta(0)+\int_{0}^{t}\Phi^{-1}(s)a(s)\left(\Omega\otimes I_{N}\right)dW(s)\right).
  \end{equation}
By reduction to absurdity, we assume $\int_{0}^{\infty}a(s)ds<\infty$. Thus, we know $\lim_{t\rightarrow\infty} \Phi(t)=C_{2}\neq0$ and 
$$\lim_{t\rightarrow\infty}\left(\delta(0)+\int_{0}^{t}\Phi^{-1}(s)a(s)\left(\Omega\otimes I_{N}\right)dW(s)\right) \neq0,$$ 
which contradicts $\lim_{t\rightarrow\infty}\delta(t)=0$ a.s.,  i.e., the multi-agent system (\ref{m-first}) with protocol (\ref{p-first}) cannot solve the containment control almost surely. Hence, we have $\int_{0}^{\infty}a(s)ds=\infty$. $\blacksquare$

\begin{remark}\label{remark-1}
When $\sigma_{ji}=0$ for any $(s_{j}, s_{i})\in \mathscr{E}$, i.e., the dynamic network degenerates to the noise-free case, the multi-agent system (\ref{m-first}) with protocol (\ref{p-first}) can be written as $\dot{\delta}(t)=-a(t)\left(\mathscr{L}_{\mathscr{F}\mathscr{F}}\otimes I_{N}\right)\delta(t)$. It is easy to know that the multi-agent system (\ref{m-first}) with protocol (\ref{p-first}) solves the containment control if and only if (A1)--(A2) hold. When $a(t)=1$, the results can be found in Theorem 1 of \cite{liu12Auto}.
\end{remark}

\section{\bf Containment control with multiple dynamic leaders}\label{s-dynamic}

In this section, we consider a multi-agent system which consists of multiple dynamic leaders. Each leader with second-order integrator dynamics is given as follows:
\begin{equation}\label{m-second}
   \left\{
   \begin{aligned}
   &\dot{x}_{i}(t)=v_{i}(t) \\
   &\dot{v}_{i}(t)=u_{i}(t) &i \in \mathscr{R}, \\
   \end{aligned}
   \right.
  \end{equation}
where $x_{i}\in \mathbb{R}^{N}$, $v_{i}\in \mathbb{R}^{N}$ and $u_{i}\in \mathbb{R}^{N}$ are the position, velocity and control input, respectively, of agent $i$. The dynamics of each follower is given in (\ref{m-first}).

We give a protocol with measurement noises as follows:
\begin{equation}\label{p-second}
   u_{i}(t)=\left\{
   \begin{aligned}
   &a(t)\left[k\sum_{j\in \mathscr{F}\cup\mathscr{R}}a_{ij}(y_{ji}(t)-x_i(t))+v_{i}(t)\right], & i \in \mathscr{F},\\
   &\dot{a}(t)\mathbf{1}_{N},   &i\in \mathscr{R},\\
   \end{aligned}
   \right.
  \end{equation}
where $y_{ji}(t)$ is defined in Section \ref{s-stationary}, $a(t): [0, \infty)\rightarrow [0, \infty)$ is a continuous differentiable function,  $\dot{v}_{i}(t)=\gamma ka(t)\sum_{j\in \mathscr{F}\cup\mathscr{R}}a_{ij}(y_{ji}(t)-x_i(t))$ for $i \in \mathscr{F}$, $k>0$, $0<\gamma<1$ are the feedback gains. Then, we have
\begin{equation}\label{m-second-p}
\left\{
   \begin{aligned}
  & dx_{\mathscr{F}}(t)=-a(t)\left(k\left(\mathscr{L}_{\mathscr{F}\mathscr{F}}\otimes I_{N}\right)x_{\mathscr{F}}(t)dt\right.\\
  &~~~~\left.+k\left(\mathscr{L}_{\mathscr{F}\mathscr{R}}\otimes I_{N}\right)x_{\mathscr{R}}(t)dt-v_{\mathscr{F}}(t)dt-k\left(\Omega\otimes I_{N}\right)dW(t)\right),\\
  &dv_{\mathscr{F}}(t)=-a(t)\left(\gamma k\left(\mathscr{L}_{\mathscr{F}\mathscr{F}}\otimes I_{N}\right)x_{\mathscr{F}}(t)dt\right.\\
  &~~~~\left.+\gamma k\left(\mathscr{L}_{\mathscr{F}\mathscr{R}}\otimes I_{N}\right)x_{\mathscr{R}}(t)dt-\gamma k\left(\Omega\otimes I_{N}\right)dW(t)\right),\\
  &dx_{\mathscr{R}}(t)=v_{\mathscr{R}}(t)dt,\\
  &v_{\mathscr{R}}(t)=a(t)C,
  \end{aligned}
   \right.
  \end{equation}
where $C\in \mathbb{R}^{mN}$ is a constant vector.

\begin{theorem}\label{T-second-mean}
Consider a directed fixed network $\mathscr{G}(\mathscr{A})$. Assume that (A3) (or A4) holds and $k>\frac{\lambda_{\max}(P)}{2\gamma(1-\gamma^{2})}$. Then, the multi-agent system (\ref{m-second-p}) solves the containment control in mean square if and only if (A1)--(A2) hold.
\end{theorem}
{\bf Proof.} Sufficiency. Because Assumption (A1) holds, we know that $\mathscr{L}_{\mathscr{F}\mathscr{F}}$ is positive definite. Let $\delta_{x}(t)=x_{\mathscr{F}}(t)+\left(\left(\mathscr{L}_{\mathscr{F}\mathscr{F}}^{-1}\mathscr{L}_{\mathscr{F}\mathscr{R}}\right)\otimes I_{N}\right)x_{\mathscr{R}}(t)$, $\delta_{v}(t)=v_{\mathscr{F}}(t)+\left(\left(\mathscr{L}_{\mathscr{F}\mathscr{F}}^{-1}\mathscr{L}_{\mathscr{F}\mathscr{R}}\right)\otimes I_{N}\right)C$. Then, we have
\begin{equation}\label{m-delta-second}
\left\{
   \begin{aligned}
  &d\delta_{x}(t)=-a(t)k\left(\mathscr{L}_{\mathscr{F}\mathscr{F}}\otimes I_{N}\right)\delta_{x}(t)dt+a(t)\delta_{v}(t)dt\\
  &~~~~~~~~~~+a(t)k\left(\Omega\otimes I_{N}\right)dW(t),\\
  &d\delta_{v}(t)=-a(t)\gamma k\left(\mathscr{L}_{\mathscr{F}\mathscr{F}}\otimes I_{N}\right)\delta_{x}(t)dt\\
  &~~~~~~~~~~+a(t)\gamma k\left(\Omega\otimes I_{N}\right)dW(t),
  \end{aligned}
   \right.
  \end{equation}
which can be rewritten as follows:
\begin{equation}\label{m-delta-second-2}
d\delta(t)=-a(t)F\delta(t)dt+a(t)\Upsilon dW(t),
  \end{equation}
where $\delta(t)=\begin{pmatrix} \delta_{x}(t)\\ \delta_{v}(t)  \end{pmatrix}$, $F=\begin{pmatrix}k\mathscr{L}_{\mathscr{F}\mathscr{F}}  & -I_{n-m} \\ \gamma k\mathscr{L}_{\mathscr{F}\mathscr{F}}& 0_{n-m} \end{pmatrix}\otimes I_{N}$ and $\Upsilon=\begin{pmatrix} k\Omega\\ \gamma k\Omega \end{pmatrix}\otimes I_{N}$.

Because $0<\gamma<1$ and $P$ is positive definite, one knows that $\begin{pmatrix}P  & -\gamma P \\ -\gamma P& P \end{pmatrix}:=\bar{P}$ is positive definite.
Due to $k>\frac{\lambda_{\max}(P)}{2\gamma(1-\gamma^{2})}$ and
\[
\begin{aligned}
  \bar{P}F+F^{T}\bar{P}=:Q= \begin{pmatrix}k(1-\gamma^{2})I_{n-m}  & -P \\ -P& 2\gamma P \end{pmatrix},
\end{aligned}
\]
it shows that $Q$ is positive definite.

Let the Lyapunov function $V(t)=\delta^{T}(t)\left(\bar{P}\otimes I_{N}\right)\delta(t)$.
Similar to the proof of Theorem \ref{T-first-mean} (or Theorem \ref{T-first-mean-2}), we know that the multi-agent system (\ref{m-second-p}) solves the containment control in the sense of mean square.

Necessity. The necessity can be proved by the same argument as in the proof of the necessity in Theorem \ref{T-first-mean}.  $\blacksquare$

Similar to the proof of Theorem \ref{T-first-almost}, we can get the following result directly.
\begin{theorem}\label{T-second-almost}
Consider a directed fixed network $\mathscr{G}(\mathscr{A})$. Assume that (A3) holds and $k>\frac{\lambda_{\max}(P)}{2\gamma(1-\gamma^{2})}$. Then, the multi-agent system (\ref{m-second-p}) solves the containment control almost surely if and only if (A1)--(A2) hold. $\blacksquare$
\end{theorem}


\section{\bf Simulations }\label{s-Simulations}

\begin{figure}
 \centering
  \includegraphics[width=5.00cm]{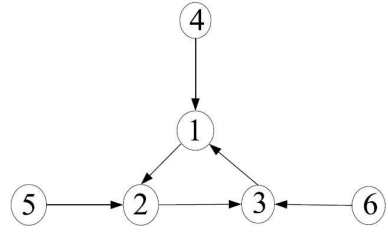}\\
  \caption{A directed network $\mathscr{G}$. }\label{fig1}
\end{figure}

\begin{figure}
 \centering
  \includegraphics[width=8.00cm]{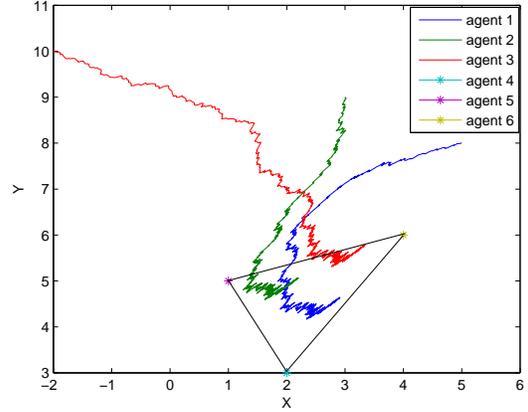}\\
  \caption{The states of agents (\ref{m-first}) with protocol (\ref{p-first}).}\label{fig2}
\end{figure}

\begin{figure}
 \centering
  \includegraphics[width=8.00cm]{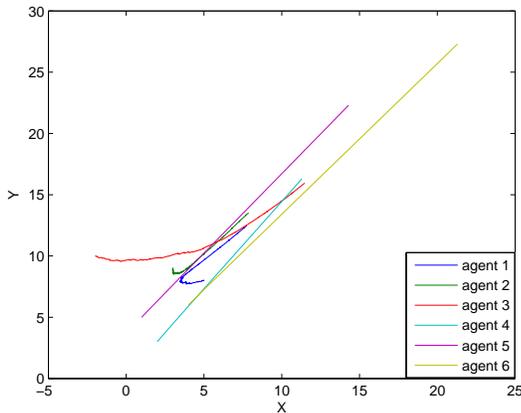}\\
  \caption{The states of multi-agent system (\ref{m-second-p}).}\label{fig3}
\end{figure}

A directed network $\mathscr{G}$  is provided in Fig. \ref{fig1}. Assume that $a_{ij}=\sigma_{ij}=1$ for $i, j \in \mathcal{I}_{n}$ if $j\in \mathscr{N}_{i}$, consensus gain function $a(t)=\frac{log(t+1)}{t+1}$. Thus, the directed network $\mathscr{G}$ has a directed spanning tree and$\int_{0}^{\infty}a(s)ds=\infty$,  $\int_{0}^{\infty}a^{2}(s)ds<\infty$. Let $N=2$. By using the protocol (\ref{p-first}), the states of agents in (\ref{m-first}) are shown in Fig. \ref{fig2}. Let $k=1$, $\gamma=0.5$, the states of agents in (\ref{m-second-p}) are shown in Fig. \ref{fig3}, which is consistent with the theoretical results.

\section{\bf Conclusions}\label{s-Conclusion}

In this paper, the containment control of multi-agent systems with measurement noises is considered. The stochastic approximation type protocols are presented for solving the containment control problem. If $\int_{0}^{\infty}a^{2}(s)ds<\infty$ (or $\lim_{t\rightarrow\infty} a(t)=0$), some necessary and sufficient conditions are obtained to ensure the followers converge to the convex hull spanned by the multiple stationary and dynamic leaders in the sense of mean square, respectively. For almost surely containment control problem, some necessary and sufficient conditions are also given by using supermartingale theory if $\int_{0}^{\infty}a^{2}(s)ds<\infty$. The future work will focus on the containment control of multi-agent systems with measurement noises under switching topologies.



\end{document}